\documentclass[10pt,a4paper,twocolumn,english,prl, showpacs, aps, superscriptaddress]{revtex4-1}
\usepackage[T1]{fontenc}
\usepackage[latin9]{inputenc}
\usepackage{units}
\usepackage{amsmath}
\usepackage{amssymb}
\usepackage{graphicx}

\makeatletter

\pdfpageheight\paperheight
\pdfpagewidth\paperwidth

 \@ifundefined{textcolor}{}
 {%
   \definecolor{BLACK}{gray}{0}
   \definecolor{WHITE}{gray}{1}
   \definecolor{RED}{rgb}{1,0,0}
   \definecolor{GREEN}{rgb}{0,1,0}
   \definecolor{BLUE}{rgb}{0,0,1}
   \definecolor{CYAN}{cmyk}{1,0,0,0}
   \definecolor{MAGENTA}{cmyk}{0,1,0,0}
   \definecolor{YELLOW}{cmyk}{0,0,1,0}
 }

\usepackage{amssymb, amsmath, color, graphicx} 

\makeatother

\usepackage{babel}
\begin{document}

\title{Characterization of S-$\mathrm{T_{+}}$ Transition Dynamics via Correlation
Measurements}

\author{Christian Dickel}
\altaffiliation{Present address: 
QuTech and Kavli Institute of Nanoscience, Delft University of Technology,
2600 GA Delft, The Netherlands
}

\affiliation{JARA-Institute for Quantum Information, RWTH Aachen University, D-52074 Aachen, Germany}

\author{Sandra Foletti}

\affiliation{Department of Physics, Harvard University, Cambridge, Massachusetts
02138, USA}

\author{Vladimir Umansky}
\affiliation{Braun Center for Submicron Research, Department of Condensed Matter
Physics, Weizmann Institute of Science, Rehovot 76100, Israel}

\author{Hendrik Bluhm}
\affiliation{JARA-Institute for Quantum Information, RWTH Aachen University, D-52074 Aachen, Germany}

\begin{abstract}
Nuclear spins are an important source of dephasing for electron spin
qubits in GaAs quantum dots. Most studies of their dynamics have
focused on the relatively slow longitudinal polarization. We show,
based on a semiclassical model and experimentally, that the dynamics
of the transverse hyperfine field can be probed by correlating
individual Landau-Zener sweeps across the S-$\mathrm{T_{+}}$
transition of a two-electron spin qubit. The relative Larmor precession
of different nuclear spin species leads to oscillations in these
correlations, whose decay arises from dephasing of the nuclei. In the
presence of spin orbit coupling, oscillations with the absolute Larmor
frequencies whose amplitude reflects the spin orbit coupling strength
are expected. 

\end{abstract}


\maketitle

Electron spins in semiconductor quantum dots have been studied much
as a platform for quantum information processing. While at first it
was envisioned to use single electron spins as qubits \citep{Loss1998},
it is also possible to encode a qubit in several electron spins. 
The experimentally most established approach is to use the singlet
state S and the $m=0$ triplet state, $\mathrm{T_{\text{0}}}$, of
two electrons in a double quantum dot (DQD). Alternatively, it was proposed
\citep{Petta2010} to replace $\mathrm{T_{\text{0}}}$ with the $m=1$
state, $\mathrm{T_{+}}$. Coherent oscillations between these two
states were observed using Landau-Zener-Stückelberg interferometry
\citep{Petta2010}, and pulse sequences to control this type of qubit
have been constructed \citep{Petta2010,Ribeiro2012b,Ribeiro2013}.
The transition between S and  $\mathrm{T_{\text{0}}}$ 
is a narrow avoided crossing mediated
by perpendicular effective magnetic fields arising from nuclear spins
and spin orbit (SO) coupling \citep{Stepanenko2012}. It is thus important
to understand the quantum dynamics of transitions between these two
states, which, as we show here, are strongly affected by the precession
of the nuclear spins in the external field. Nuclear spin effects are
especially prominent in GaAs-based quantum dots, currently the most
developed material system available for gated semiconductor quantum
dots.

In addition to qubit control, the S-$\mathrm{T_{\text{+}}}$ transition
can be used to transfer angular momentum from the electronic system to
the nuclear spins. The resulting dynamic nuclear polarization (DNP)
has been instrumental in reducing dephasing \citep{Bluhm2010a} and
achieving universal control \citep{Foletti2009} of
S-$\mathrm{T_{\text{0}}}$ qubits. Periodically traversing the crossing
can be used to polarize the nuclear spins and even to suppress
fluctuations via a feedback loop. Such nuclear spin control is useful
for both gate defined and self assembled quantum dots
\citep{Kloeffel2011,Latta2009,Ladd2010}.  In gated dots, the
performance of such DNP schemes is currently limited by a low spin
transfer probability from electrons to nuclei \citep{Bluhm2010a},
which is poorly understood. It was observed that the electron spin
flip rate exceeds the nuclear polarization rate, which may result from
an additional SO spin flip channel
\citep{Rudner2010,Brataas2011,Brataas2012,Neder2014}.  Probing the transitions
between S and $\mathrm{T_{+}}$ states and the role of SO coupling is
thus also relevant for understanding the limitations of DNP methods.

Here, we present a measurement technique that probes the dynamics
of the matrix element driving transitions between S and $\mathrm{T_{+}}$.
Direct measurements of the Overhauser field have previously been used
to probe fluctuations of its component parallel to the external magnetic
field \citep{Barthel2009,Bluhm2010a}. However, the transverse part
that drives the oscillations between S and $\mathrm{T_{+}}$ exhibits
faster dynamics that are harder to access. Our approach is based on
correlating the outcomes of subsequent individual single shot measurements
of the qubit state after Landau-Zener (LZ) transitions similar to
\citep{Fink2013a}. After each LZ sweep, the state of the qubit is
measured. From this data, the auto-correlation of single shot measurements
is computed as a function of the time delay between them. The time
resolution is only limited by the accuracy of pulse timing, but not
by the required averaging time. 

We will show in the following that these correlations reveal the
relative nuclear spin Larmor precession and dephasing of these
precessions on a $\mu$s scale. In the presence of SO coupling, the
oscillations would exhibit additional frequency components. Therefore,
the method can provide direct, qualitative and quantitative evidence
for SO coupling.  We also present proof of principle experimental
results. They show nuclear spin effects but no SO coupling. We also
point out that averaging the  LZ transition probability over 
nuclear spin configurations leads to an algebraic instead of exponential 
dependence on the sweep rate. A SO component of the transition matrix
element is not subject to this averaging and accordingly leads to more
exponential behavior.

\begin{figure}
\begin{centering}
\includegraphics[width=1\columnwidth]{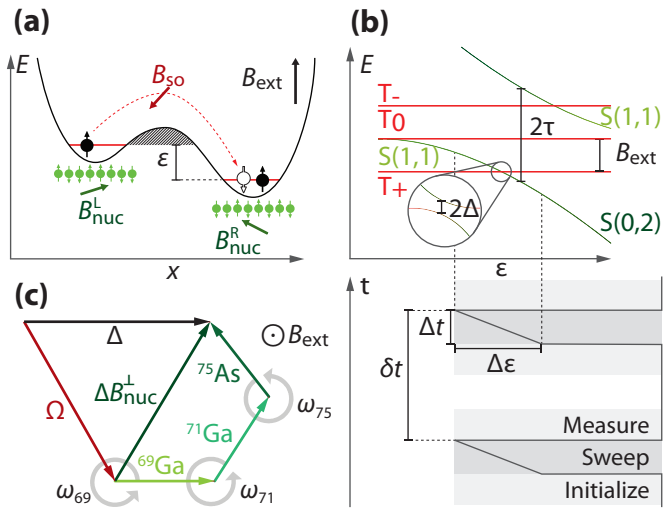}
\par\end{centering}

\caption{(a) DQD potential landscape along the dot-dot axis. Vectors indicate
effective magnetic fields from SO due to tunneling and nuclear spins
via the Fermi contact interaction. The detuning $\varepsilon$ is
the energy difference between the single electron levels in the dots. (b)
DQD energy diagram and pulse forms for LZ correlation measurements.
Pulse cycles are concatenated with a time delay $\delta t$. (c) $\boldsymbol{\Delta B}_{\mathrm{\mathrm{nuc}}}^{\perp}$
has contributions from the three different nuclear species that precess
in $B_{\mathrm{ext}}$ with different Larmor frequencies. The S-$\textrm{T}_{+}$transition
matrix element $\Delta$ is a vector sum of the static SO
field $\boldsymbol{\Omega}$ and $\boldsymbol{\Delta B}_{\mathrm{\mathrm{nuc}}}^{\perp}$\label{fig:1} }
\end{figure}

The two-electron spin qubits considered here consist of two electrons trapped
in a double-well potential as seen in Fig. \ref{fig:1}(a) formed
by electrostatic gating of a two-dimensional electron gas in a GaAs/AlGaAs
heterostructure. A comprehensive derivation of the relevant DQD electronic
Hamiltonian can be found in \citep{Stepanenko2012}. In the following
we provide a qualitative summary. The qubit is controlled via the
detuning, $\varepsilon$, defined as the energy difference between the lowest
(1, 1) and (0, 2) charge configuration, where the numbers indicate
the occupancy of each dot. Positive detuning favors double occupation
of one dot in the singlet state, while at negative detuning the electrons
are separated. The resulting level diagram as a function of $\varepsilon$
is shown in Fig. \ref{fig:1}(b). The singlet states exhibit an
avoided crossing with tunnel coupling $\tau$. Due to the Pauli principle,
the triplet states $\vert\mathrm{T_{+}}\rangle$, $\vert\mathrm{T_{-}}\rangle$
and $\vert\mathrm{T_{\text{0}}}\rangle$ with magnetic quantum numbers
$m=\pm1,0$ always stay in the (1, 1) configuration and are energetically
separated by the Zeeman energy in an external magnetic field $\mathit{B}_{ext}$.
This setting allows fast electric singlet initialization and single
shot readout that distinguishes singlet and triplet states via spin
to charge conversion \citep{Barthel2009}, independent of
which triplet is used.

The focus of this Letter is the S-$\textrm{T}_{+}$ degeneracy point
(called S-$\textrm{T}_{+}$ transition), where transitions between
these states are possible. A convenient way to probe this narrow crossing
is to employ LZ sweeps, i.e., linear sweeps across the transition as illustrated
in Fig. \ref{fig:1}(b) that do not require precise knowledge of
the position of the transition. For a sufficiently small sweep range,
the other states can be neglected. We use an effective 
Hamiltonian in the basis $\ensuremath{\left\{ \vert\mathrm{S^{*}\left(\varepsilon\right)\rangle,\vert T_{+}}\rangle\right\} }$,
where $\vert\mathrm{S}^{*}\left(\varepsilon\right)\rangle=\cos\psi\vert\mathrm{S\left(1,1\right)}\rangle+\sin\psi\vert\mathrm{S\left(0,2\right)}\rangle$
is the hybridized singlet close to the crossing:

\begin{equation}
\mathbf{H}=\mathrm{\frac{\mathit{B}_{ext}-\mathit{J}\left(\varepsilon\right)}{2}}\sigma_{z}+\Delta\sigma_{x},\label{eq:Landau-Zener Hamiltonian}
\end{equation}

where $\sigma_{x}$ and $\sigma_{z}$ are Pauli
matrices. We use units of energy for all electronic Zeeman fields,
thus setting $g\mu_{B}=1$. The parameters are the exchange splitting
$J=E_{S^{*}}-E_{T_{0}}$ and the coupling 
\begin{equation}
\Delta=\frac{1}{2}\left|\boldsymbol{B}_{\mathrm{SO}}\sin\psi+\boldsymbol{\Delta B}_{\mathrm{\mathrm{nuc}}}^{\perp}\cos\psi\right|.\label{eq:Delta}
\end{equation}
The singlet mixing angle, $\psi$, at the transition sets
the weight of the effective SO field $\boldsymbol{B}_{\mathrm{SO}}$,
which couples $\vert\mathrm{S\left(0,2\right)}\rangle$ and
$\vert\mathrm{T_{+}}\rangle$, and the difference in transverse
hyperfine fields $\boldsymbol{\Delta
    B}_{\mathrm{\mathrm{nuc}}}^{\perp}=\left(\boldsymbol{B}_{\mathrm{\mathrm{nuc}}}^{\perp,R}-\boldsymbol{B}_{\mathrm{\mathrm{nuc}}}^{\perp,L}\right)$
between the dots, which couples
$\vert\mathrm{S\left(1,1\right)}\rangle$ and
$\vert\mathrm{T_{+}}\rangle$ \citep{Stepanenko2012}. The mixing angle
depends on $\tau$ and indirectly on $\mathit{B}_{ext}$ via the
position of the transition. For $J=\alpha t$, where $\alpha$ is the
sweep rate and a sweep from $-\infty$ to $\infty$, the probability
that an initial singlet stays in a singlet is given by
$P_{S}=\exp\left(-\frac{2\pi\Delta^{2}}{\hbar\alpha}\right)$
\citep{LevLanda1932,Zener1932}.

In this work, we assume $B_{ext} \ll \tau$ so that $\cos\psi\approx
1$ and $\sin\psi \ll 1$ at the transition. $B_{\mathrm{SO}} = \frac {4 \tau} 3
\frac{l}{l_{SO}}$ depends the inter-dot distance $l \approx 200$ nm
and the SO length $l_{SO}$ \citep{Stepanenko2012}. 
Therefore the effective field from SO coupling is
expected to be static with a value on the order of $B_{SO}\sim$ 100
mT.  For two million unit cells in each dot \citep{Bluhm2010}, 
 $\left\langle
  \left(\boldsymbol{\Delta
    B}_{\mathrm{nuc}}^{\perp}\right)^{2}\right\rangle
  ^{\nicefrac{1}{2}}$ is on the order of 4 mT. Thus even a suppressed
SO coupling could approach the expected Overhauser field. We introduce
the overall SO contribution to $\Delta$, 
$\boldsymbol{\Omega}=\boldsymbol{B}_{\mathrm{SO}}\sin\psi$. In the
regime of interest here, $\boldsymbol{\Omega} \lesssim  \left\langle
  \left(\boldsymbol{\Delta
    B}_{\mathrm{nuc}}^{\perp}\right)^{2}\right\rangle
  ^{\nicefrac{1}{2}}$.

The dynamical effects considered here originate in the time dependence
of $\Delta$ via $\boldsymbol{\Delta B}_{\mathrm{nuc}}^{\perp}$,
which arises from the interactions of each electron spin with on the
order of $10^{6}$ nuclear spins. As illustrated in Fig. \ref{fig:1}(c), 
$\boldsymbol{\Delta B}_{\mathrm{\mathrm{nuc}}}^{\perp}$ is a
vector sum of contributions from the three isotopes present in GaAs,
$^{69}$Ga, $^{71}$Ga and $^{75}$As, all with nuclear spin $\nicefrac{3}{2}$.
Following earlier work, we treat their fields as classical random
variables \citep{Neder2011}.

In addition to the Larmor precession of the nuclei, we also account
for dephasing of this precession due to microscopic field fluctuations
originating from dipole-dipole interactions and quadrupole splittings.
To this end, we subdivide each species into subspecies with slightly
different precession frequencies $\omega_{i}=\gamma_{i}\left(B_{\mathrm{ext}}+\delta B_{i}\right)$,
where $\gamma_{i}$ is the nuclear gyromagnetic ratio. The field fluctuations
$\delta B_{i}$ are taken from a discretized normal distribution with
standard deviation $\delta B_{\mathrm{loc}}$, and the index $i$
denotes both the nuclear species and the local field value. We find
that good convergence is obtained for five bins per species.

We incorporate the above dynamics by writing
\begin{eqnarray}
\Delta\left(t\right)^{2} & = &\frac 1 4 \Bigl(\sum_{ij}B_{i}^{+}B_{j}^{-}e^{i\left(\omega_{i}-\omega_{j}\right)t}\nonumber \\
 &  & +\sum_{i}\Omega\left(B_{i}^{+}e^{i\omega_{i}t}+B_{i}^{-}e^{-i\omega_{i}t}\right)+\Omega^{2}\Bigr),\label{eq:Delta Formel ausgeschrieben}
\end{eqnarray}
where $B_{i}^{\pm}=B_{i}^{x}\pm iB_{i}^{y}$. Note that the contribution
from SO coupling behaves like an additional, static nuclear spin species.
During a single LZ sweep, we approximate $\Delta$ as constant because
the nuclear spin dynamics are slow compared to the time spent at the
crossing. This approximation is justified as long as $\gamma_{N}/\gamma_{e}B_{ext}\ll\Delta$,
where $\gamma_{N}$ and $\gamma_{e}$ are nuclear and electronic gyromagnetic
ratios. 

The expectation values over nuclear states of the single shot measurement
correlations are given by $\left\langle P_{S}\left(t\right)P_{S}\left(t+\delta t\right)\right\rangle =\left\langle \exp\left(-\frac{2\pi}{\alpha\hbar}(\Delta\left(t\right)^{2}+\Delta\left(t+\delta t\right)^{2}\right)\right\rangle $ \citep{Fink2013a}.
To evaluate them one has to integrate over all possible initial conditions
$B_{i}^{\pm}$. In the homogeneous coupling approximation used here, each
component of the effective field of each species is approximated as
normally distributed with standard deviation $B_{i, \mathrm{rms}}=\frac{A_{i}}{N}\sqrt{5 N_{i}/2}$,
which corresponds to averaging over an infinite temperature nuclear
ensemble. $N_{i}$ is the number of spins in the $i$th bin, $A_{i}$
the species-dependent hyperfine coupling strength and $N$ the number of unit cells occupied by
each electron.

To perform this integration, it is
useful to decouple the terms via the T-matrix method 
\citep{Neder2011}. The exponent of the LZ correlations
is rewritten as
\begin{align*}
\frac{2\pi}{\hbar\alpha}\left(\Delta\left(t\right)^{2}+\Delta\left(t+\delta t\right)^{2}\right) & = \frac{\pi}{\alpha \hbar} \Omega^{2}+\sum_{ij}T_{ij}\left(\delta t\right)z_{i}^{+}z_{j}^{-}\\
 & +\sum_{i}V_{i}\left(\delta t\right)^{*}z_{i}^{-}+\sum_{i}V_{i}\left(\delta t\right)z_{i}^{+}
\end{align*}
with $z_{i}=x_{i}+ i y_{i}$, where $x$ and $y$ are normally
distributed with unit variance. Diagonalizing the T-matrix, applying
the same basis change to the linear terms and integrating over the
$z_{i}$ distributions results in
\begin{eqnarray*}
\langle P_{S}\left(t\right)P_{S}\left(t+\delta t\right)\rangle &
= & \exp\left(- \frac{\pi \Omega^2}{\alpha \hbar} \right) \\
&\times&
\prod_{i}\frac{1}{1+2\lambda_{i}}\cdot\exp\left(\frac{2 \tilde V_{i}{}^{2}}{\left(1+2\lambda_{i}\right)}\right),
\end{eqnarray*}
where $\lambda_{i}$ are the eigenvalues of the T-matrix
and the $\tilde V_{i}$ are the components of the vector $V_{i}$
in the eigenbasis of the T-matrix. A similar, straightforward
calculation also reveals a modification of the average flip probability
due to averaging over nuclear spin configurations: 
\begin{equation}
\left\langle P_{S}\right\rangle 
= \frac{\alpha\hbar}{\alpha\hbar + \pi \sum_{i} B_{i,\mathrm{rms}}^{2}}
\cdot\exp\left(-\frac{\pi\Omega^2}{2\left(\alpha\hbar + \pi \sum_{i} B_{i,\mathrm{rms}}^{2}\right)}\right).\label{eq:averaged-LZ}
\end{equation}
For $\Omega=0$, the behavior changes from exponential to algebraic.

\begin{figure}
\begin{centering}
\includegraphics[width=1\columnwidth]{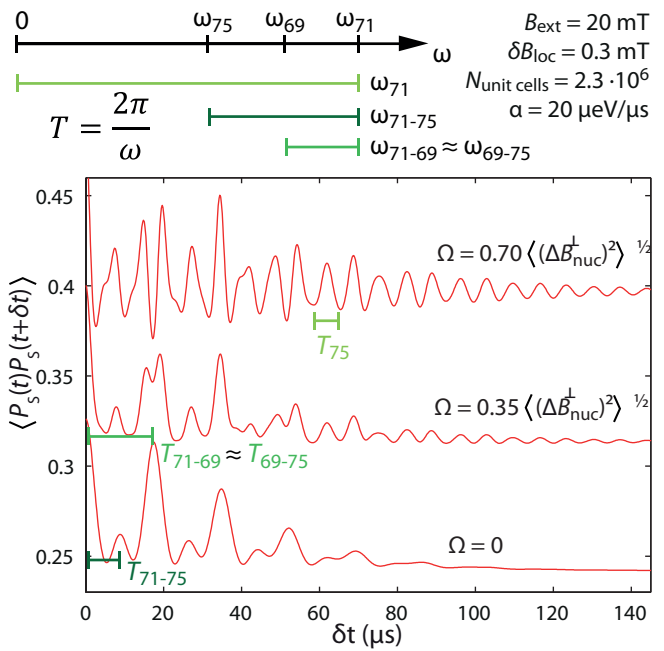}
\par\end{centering}

\caption{Simulations of correlations for different SO coupling.  Above
  the nuclear spin frequencies are given. For $\Omega=0$, correlations
  oscillate with the relative Larmor frequencies. For $\Omega>0$
  additional frequencies appear, and the corresponding oscillations
  extend to longer times while the original peaks become smaller and
  some split into two.\label{fig:2}}
\end{figure}

Fig. \ref{fig:2} shows the correlations for typical experimental
parameters and different SO coupling constants $\Omega=0$. As also
seen in equation \ref{eq:Delta Formel ausgeschrieben}, $\Delta$
oscillates with the Larmor frequency differences $\omega_{i}-\omega_{j}$
for $\Omega=0$. In GaAs, the three $\gamma_{i}$ are approximately
equidistant. Therefore, the relative orientations of the 
three  contributions to $\boldsymbol{\Delta B}_{\mathrm{\mathrm{nuc}}}^{\perp}$ 
and thus $\Delta$ return to nearly the same value after
one period $T_{ij}=2\pi\left(\omega_{i}-\omega_{j}\right)^{-1}$ of the
slower relative precession, $T_{75-69}\approx T_{71-69}$. 
Hence, large revivals in the correlations occur at this value of $\delta t$.
The smaller peaks at $\delta t=T_{75-71}$ result from a partial realignment
of these two species only. The envelope decay of the correlations
on a time scale of approximately 40 $\mu$s arises from nuclear spin
dephasing. 

As $\Omega$ is increased, oscillations at the bare frequencies
$\omega_{i}$ extending to larger $\delta t$ start to appear as well.
In addition, the large peak around $\delta t=17\:\mu\mathrm{s}$ splits.
Observing these features would represent qualitative evidence for
SO coupling. At the same time, the magnitude of the variation in the
correlations is reduced.

We now turn to proof of principle experimental results, obtained from
the same DQD sample as used in \citep{Bluhm2010}. In all measurements,
the sweep range was constrained to keep $J\left(\varepsilon\right)\gg B_{nuc}$
to prevent mixing of S and T$_{0}$. As $B_{ext}\gg\Delta$, the nonlinear
sweep and the finite sweep range do not introduce a large error. A
measured background reflecting the direct coupling of the gate pulses
to the charge sensor used for single shot readout was subtracted from
each dataset.

\begin{figure}
\begin{centering}
\includegraphics[width=1\columnwidth]{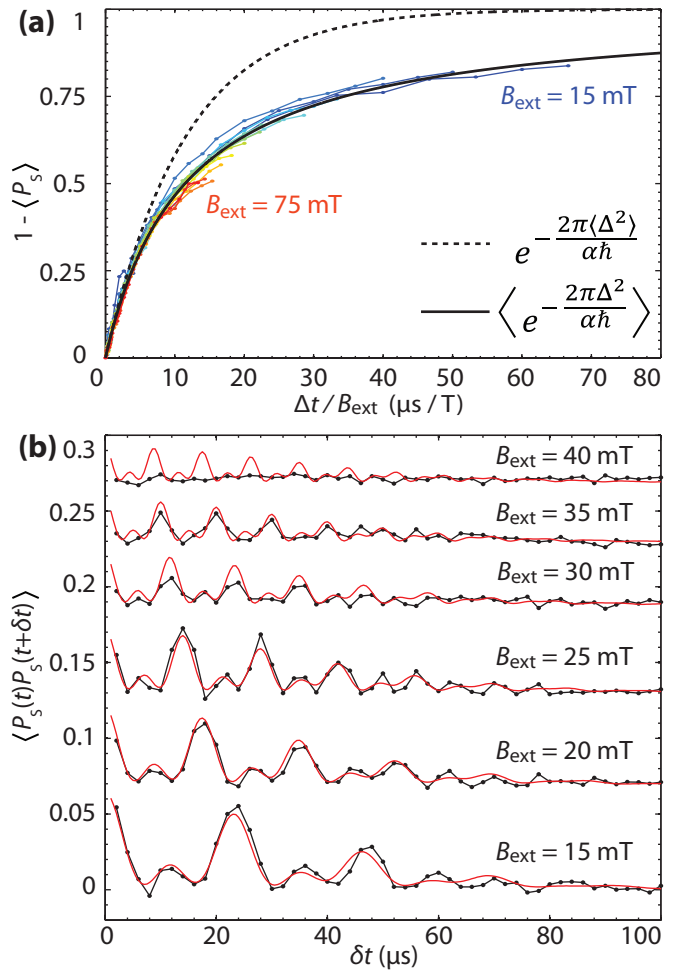}
\par\end{centering}

\caption{(a) Triplet probabilities as a function of sweep-time $\Delta t$
from 0 $\mu$s to 1 $\mu$s with a constant $\Delta\varepsilon$ for
different $B_{\mathrm{ext}}$. The scaling of the 
time axis with magnetic field leads to a reasonable collapse of the curves. Data
in color. The  black lines shows the LZ-model (dashed) with a fixed transition 
matrix element, and the nuclear spin averaged LZ behavior (solid).
With $N=2.3\cdot10^{6}$ fixed, $\Delta^2/\alpha$ requires a scaling factor
of 6 to match experiment. (b) Correlation of triplet probabilities
for different $B_{\mathrm{ext}}$, data in black and model in red.
Except for the magnitude, the correlations behave as expected as a function 
of $B_{\mathrm{ext}}$.
\label{fig:3}}
\end{figure}

To verify the expected behavior of flip the probability, we measured its dependence
on sweep rate and magnetic field, as shown in Fig. \ref{fig:3}(a).
The scaling factor used to convert the readout signal to triplet probability
was chosen in accordance with the contrast of the $(1, 1)$-$(0, 2)$ charge
transition, taking relaxation during the readout stage into account.
The scaling of the $x$-axis is motivated by the fact that in the experiment
the sweeps result in a linear time dependence of $\varepsilon$ rather
than $J\left(\varepsilon\right)$. To compute the sweep rate $\alpha=\nicefrac{dJ}{dt}=\nicefrac{dJ}{d\varepsilon}\cdot\nicefrac{d\varepsilon}{dt}$,
we employ the phenomenological relation $J\left(\varepsilon\right)=J_{0}\exp\left(-\frac{\varepsilon}{\varepsilon_{0}}\right)$
\citep{Dial2013}. At the S-T+ transition, $J\left(\varepsilon\right)=B_{\mathrm{ext}}$,
so that $\alpha=B_{\mathrm{ext}}\frac{1}{\varepsilon_{0}}\frac{\Delta\varepsilon}{\Delta t}$,
where $\Delta t$ is time taken to sweep detuning by $\Delta\varepsilon$.
Hence, $P_{S}$ is expected to be a function of $\Delta t/B_{\mathrm{ext}}$,
independent of $B_{\mathrm{ext}}$. Indeed, the curves for different $B_{\mathrm{ext}}$ 
in Fig. \ref{fig:3}(a) collapse reasonably well onto each other. 
The deviations for slow sweeps may be a result of relaxation. 
The solid line was obtained from Eq. \ref{eq:averaged-LZ}  using $\Omega=0$ and
a value of $\Delta^{2}/\alpha$ that is a factor 6 smaller than expected
based on measurements of $T_{2}^{*}$, from which $N=2.3\cdot10^{6}$
can be extracted \citep{Bluhm2010}. 
This adjustment was required to obtain adequate
agreement with the model, which contains no further free parameters.
As can be seen from the relatively large slope at large $\Delta t$,
the algebraic dependence obtained from nuclear averaging in equation
\ref{eq:averaged-LZ} gives a better overall fit than the exponential LZ formula
(dashed line), even if adjusting the scaling factor of the latter.

Guided by these results, we fixed $\Delta t$ at 0.3 $\mu$s for the
correlation measurements discussed next. Initialization--LZ-sweep--readout
cycles as shown in Fig. \ref{fig:1}(b) were repeated at a rate of
0.5 MHz. Initialization was carried out via a standard procedure through
electron exchange with the leads \citep{Petta2010}. Readout was accomplished
via RF reflectometry of a quantum point contact \citep{Barthel2009}.
Instead of single shot discrimination, we computed the correlations directly 
from the readout signal, which was averaged over 1 $\mu\mathrm{s}$
in each cycle. Assuming the readout signal is the sum of a contribution
from the qubit measurement and noise uncorrelated with the qubit state,
this procedure is equivalent to correlating single shot readout values
apart from additional noise. Due to the fixed sampling rate, the time
resolution was limited to 2 $\mu\mathrm{s}$.

The data shown in Fig. \ref{fig:3}(b) clearly shows the expected
oscillations. For fitting our model to the experiment, only an offset
of the QPC signal was adjusted individually for each data set. The
$y$-scale was set analogous to Fig. \ref{fig:3}(a). 
Obtaining good agreement of the magnitude required
$\Delta^{2} / \alpha$ to be chosen a factor 18 smaller
than expected for $N=2.3\cdot10^{6}$ unit cells and experimental sweep
rates. The $\delta t$-dependence is unaffected by this adjustment.
Nuclear spin Larmor frequencies were taken from \citep{Condensee1977}.
$\delta B_{\mathrm{loc}}$ determines the time scale of the overall
decay and was fixed to 0.33 mT, consistent with earlier measurements
\citep{Bluhm2010}. As expected for the low magnetic fields considered
here, the effect of SO coupling is negligible. At higher fields, contrast
is lost faster than expected from our model, but the time resolution
used here also becomes too coarse to resolve the Larmor precession.

The experiments thus confirm the overall picture obtained from our model
and the usefulness of single shot correlation measurements.
The significant quantitative discrepancy in $\Delta^{2} / \alpha$
is not understood, but may at least partly arise from noise and non-uniformity
of the sweep resulting from the discrete 1 ns sampling interval of
the wave form generator. More detailed measurements would be required
to explore these effects. By varying the delay between subsequent pulses
instead of using a fixed repetition rate, the time resolution could be 
improved significantly. This would enable an extension to larger magnetic 
fields where SO coupling might become important. 

In conclusion, we have shown that correlation measurements of
Landau-Zener sweeps through the S-T+ transition reveal nuclear spin
precession and dephasing. These dynamics
are seen in an experimental implementation of the scheme. They
imply that the perpendicular component of the Overhauser
field cannot be controlled as well as the $z$-component. Our
model indicates that the same measurement technique should be able to
qualitatively reveal SO coupling at higher magnetic fields.

\begin{acknowledgments}
This work was supported by the Alfried Krupp von Bohlen und Halbach
Foundation and DFG under Grant No. BL 1197/2-1. We would like to thank
Amir Yacoby, Mark Rudner and Izhar Neder for discussions.
\end{acknowledgments}

\bibliographystyle{apsrev4-1}

\end{document}